\documentclass[a4paper, 12pt]{article}
\usepackage[cmex10]{amsmath}
\usepackage{graphicx,amssymb,amsmath,amsthm,bm,amsfonts,amsmath}

\begin{document}
\title{Price and capacity competition in balancing markets with energy storage}
\author{
Joshua A. Taylor\footnote{Electrical and Computer Engineering, University of Toronto, Toronto, ON M5S 3G4, Canada, \texttt{josh.taylor@utoronto.ca}}, Johanna L. Mathieu\footnote{Electrical Engineering and Computer Science, University of Michigan, Ann Arbor, MI 48109, USA, \texttt{jlmath@umich.edu}}, Duncan S. Callaway\footnote{Energy and Resources Group, University of California, Berkeley, CA 94720, USA, \texttt{dcal@berkeley.edu}},\\and Kameshwar Poolla\footnote{Mechanical Engineering, University of California, Berkeley, CA 94720, USA, \texttt{poolla@berkeley.edu}}
}
\maketitle

\newtheorem{theorem}{Theorem}
\newtheorem{lemma}{Lemma}

\begin{abstract}
Energy storage can absorb variability from the rising number of wind and solar power producers. Storage is different from the conventional generators that have traditionally balanced supply and demand on fast time scales due to its hard energy capacity constraints, dynamic coupling, and low marginal costs. These differences are leading system operators to propose new mechanisms for enabling storage to participate in reserve and real-time energy markets. The persistence of market power and gaming in electricity markets suggests that these changes will expose new vulnerabilities.

We develop a new model of strategic behavior among storages in energy balancing markets. Our model is a two-stage game that generalizes a classic model of capacity followed by Bertrand-Edgeworth price competition by explicitly modeling storage dynamics and uncertainty in the pricing stage. By applying the model to balancing markets with storage, we are able to compare capacity and energy-based pricing schemes and to analyze the dynamic effects of the market horizon and energy losses due to leakage. Our first key finding is that capacity pricing leads to higher prices and higher capacity commitments, and that energy pricing leads to lower, randomized prices and lower capacity commitments. Second, we find that a longer market horizon and higher physical efficiencies lead to lower prices by inducing the storage to compete to have their states of charge cycled more frequently.
\end{abstract}


\section{Introduction}\label{intro}
High penetrations of variable renewable energy resources, such as wind and solar, call for additional regulation and load following capabilities to maintain system reliability under increased forecast errors \cite{makarov09,halamay2011reserve,oren2011wind}. Energy storage and demand response are widely considered promising solutions to renewable variability \cite{infield2004storage,carrasco2006pes,callaway11EL,Castillo2014survey}. To facilitate increased procurement and utilization of these resources, Independent System Operators (ISOs) have begun to propose new mechanisms enabling storage to participate in reserve and real-time energy markets, for example in California \cite{REMdraft}, Texas \cite{beaconenergy2011}, and New York \cite{NYISO2011}. In this paper, we refer to reserve and real-time energy markets as \textit{balancing markets}.

A variety of storage technologies are now available to power systems, such as pumped hydro, batteries, and flywheels \cite{Ibrahim2008char,Castillo2014survey}. Load aggregations that actively participate in power system operations through load shifting programs can also be approximated as virtual energy storage \cite{mathieu_energy_2013,taylor2013FlexCDC}. We will henceforth refer to both physical energy storage and active load aggregations simply as \textit{storage}. All types of storage share the following first-order characteristics:
\begin{itemize}
\item hard energy capacity constraints,
\item dynamic coupling of energy states, and
\item low marginal costs.
\end{itemize}
The hard energy capacity constraints arise from the physical limitations of each technology, e.g., reservoir size and flywheel material strength. These constraints couple the range of possible energy exchanges in one time period to the amount injected or extracted in the preceding periods. Since storage does not use fuel, its costs are limited to operation, maintenance, and initial procurement, resulting in low marginal costs \cite{peterson10economics,su09quantifying,alvarez04assessment}. These features distinguish storage from conventional generators. We discuss our modeling of storage and markets in detail in Section~\ref{storageback}.

The increasing role and distinct characteristics of storage are changing the physical and economic nature of power systems. As electricity markets are modified to accommodate storage, new vulnerabilities to market power and gaming will surface. Indeed, the persistence of events like the California Electricity Crisis \cite{Joskow2001crisis} and the more recent market manipulation by J.~P.~Morgan \cite{jpmorgan2013ferc} indicates that system operators must always be alert for strategic behavior. Based on its past successes in diagnosing market power in existing electricity markets, we believe game theory is a natural approach to assessing the vulnerabilities and guiding the design of new balancing markets with energy storage participants. A notable recent effort is \cite{Sioshansi2014when}, which analyzes strategic behavior when storage arbitrages energy between on- and off-peak times.

In this paper, we focus on storage whose purpose is to preserve system stability by providing reserves and regulation in balancing markets. Our main contributions are a new game-theoretic model of strategic behavior that captures the three aforementioned storage characteristics, and concrete design insights obtained from its application to balancing markets. Our proposed model is an extension of capacity followed by price competition, a two-stage game in which Bertrand-Edgeworth price competition is preceded by a capacity-setting stage \cite{kreps1983cournot,Acemoglu2009price}. Bertrand-Edgeworth competition \cite{bertrand1883,edgeworth1925papers,vives2001oligopoly} is a special case of supply function competition \cite{klemperer1989supply}, which has been widely applied to electricity markets because it captures generators' nonlinear fuel curves \cite{Bolle1992supply,green1992british,hobbs2002supply,baldick2004lin}. Standard Bertrand-Edgeworth competition is similarly well-suited to modeling storage because it captures its hard capacity constraint and low marginal costs.

In Section~\ref{dynprice}, we develop an extension of Bertrand-Edgeworth competition in the duopoly case using the theoretical approach of \cite{Acemoglu2009price} (which considers a deterministic demand in a single time period). Our extension is theoretically novel in its description of random energy imbalances, multiple time periods, and the dynamic coupling of storage's energy states. The resulting generalization of Bertrand-Edgeworth competition is still far more analytically tractable than general supply function competition in that it admits a complete characterization of the pricing equilibrium. Using the pricing equilibrium characterization, we subsequently develop a novel characterization of the scenario where storage first commits a capacity that then parametrizes the ensuing pricing game. All proofs are collected in the appendix. 

In Section~\ref{AEM}, we apply our results on price and capacity competition to balancing markets and obtain several new insights. We first use price competition to analyze the dynamic effects of market horizon and losses due to energy leakage. We find that if the market horizon is large and leakage small enough, firms will compete to be cycled more frequently by reducing their prices, hence improving economic competitiveness. We then use capacity competition to compare markets in which firms are paid for capacity (as in traditional reserve markets) and for energy (as in traditional real-time energy markets). Our results show that, similar to what has been observed in generation markets \cite{oren2000cap,Cramton2005cap}, firms commit more capacity and set higher prices under capacity pricing and commit less capacity and set lower, random prices under energy pricing. In Section~\ref{conc}, we state our design insights for energy balancing markets with storage and identify some future research directions.

\section{Storage in balancing markets}\label{storageback}
In this section, we describe our model of storage and the role of storage in balancing markets.

\subsection{Physical modeling}\label{storageback1}
We employ the following generic storage model (see, e.g., \cite{taylor2011store}). Let $S^t_i$ be firm $i$'s stored energy at time $t$, also known as the \textit{state of charge}, and let $X^t_i$ be the energy added or removed from the storage at time $t$. Between time periods, a fraction of the state of charge, $1-\alpha_i\in[0,1]$, is dissipated, which we refer to as \textit{leakage}. The state of charge of storage $i$ then evolves according to the difference equation
\[
S^{t+1}_i=\alpha_i S^t_i +X^t_i.
\]
The state of charge must remain within the hard capacity limit $0\leq S^t_i\leq S_i$, where $S_i$ is the maximum amount of energy that can be stored. Here we have neglected injection and extraction losses for modeling tractability; we heuristically let leakage act as a proxy for these inefficiencies because all have the effect of dissipating energy and hence reducing the state of charge.

We also neglect constraints on $X^t_i$ (analogous to a generator's power constraint) and the derivative of $X^t_i$ (analogous to a generator's ramp constraint) to maintain the tractability of our model. Ignoring power constraints is valid for storage technologies with high power and low energy capacities like flywheels and superconducting magnetic energy storage \cite{Ibrahim2008char}, which are most appropriate for the balancing markets we consider. Ignoring ramp constraints is justified because storage is generally able to ramp quickly. For comparison, coal, nuclear, and steam turbines can change their power output at 1--5\% per minute, and gas turbines and diesel generators can ramp at 5--40\% per minute \cite{vittal_impact_2009}. In contrast, a pumped hydro storage facility in Wales, UK can move from zero to full output in less than 16 seconds \cite{first_dinorwig_2012}, and newer technologies like grid-level batteries and flywheels can respond even faster \cite{Ibrahim2008char,Castillo2014survey}.

In our balancing market model, an energy imbalance $B^t$ is realized in each time period. The objective of the system operator is to eliminate the imbalance by apportioning it over a group of storages; this objective contrasts other uses of storage like load-shifting, in which useful energy is moved in time to achieve greater efficiency \cite{Castillo2014survey}. The portion of $B^t$ that can be allocated to a particular storage is limited by its state of charge if $B^t<0$ and its unused capacity, $S_i-S^t_i$, if $B^t>0$. A participating storage can be paid for the capacity it commits to the balancing market, $S_i$, or for the imbalances it absorbs, $X^t_i$. In the latter case, storage is paid a positive amount for canceling imbalances regardless of whether $B^t$ is positive or negative. This is realistic because both positive and negative imbalances are undesirable. For instance, both can result from renewable intermittency or faults and both can cause frequency instability \cite{kundur1994stab,WoodWoll2}. Also, the mileage payments some system operators have implemented in response to the Federal Energy Regulatory Commission's Order \#755 are similarly based on a price times an absolute value, albeit for ramping rather than energy or power \cite{caiso_pay_2012}.

In the special case in which there is one time period and $B$ is deterministic, we interpret $B$ as a demand for storage capacity. We make use of this special case in Section~\ref{capvsen} in comparing energy-based and capacity-based payment schemes.

Because storage does not have a fuel source, its only costs are associated with operation, maintenance, and initial procurement. Over long periods of time, these are small relative to the fuel costs incurred by generators \cite{peterson10economics,su09quantifying, alvarez04assessment}. For this reason and for concision, we model storage as having zero marginal costs associated with energy injections and extractions. Consequently, all prices higher than what would be necessary to recoup opportunity costs in our subsequent model are the result of gaming, which may be interpreted as undesirable additions `superimposed' on top of marginal cost prices from a non-strategic (e.g., welfare maximizing) model. In this regard, our approach is intended to isolate strategic behaviors, and does not account for other intended revenue streams of storage. 

Storage can provide a handful of services besides balancing such as load-shifting, contingency reserves, and voltage regulation \cite{he2011novel,sioshansia2012market,Castillo2014survey}. Here, we only consider its participation in balancing markets, and model the foregone profits from providing other services as the \textit{opportunity cost} $-\gamma_i S_i$, where $\gamma_i\geq0$ is a known parameter and $S_i$ is the capacity committed to the balancing market. Because storage cannot sell power in baseload energy markets (because it must first take in all energy that it provides), these opportunities costs may be substantially lower than those of generators participating in reserve markets \cite{WoodWoll2,Litvinov2006ex}.

\subsection{Market modeling}\label{storageback2}
As discussed in Section~\ref{intro}, balancing markets match supply with demand on faster time scales than baseload energy markets. For example, many ISOs operate real-time energy markets that balance supply and demand every five minutes, and also operate frequency regulation markets that procure capacity to balance supply and demand on timescales of seconds. If the associated baseload energy market is well designed, the imbalances in balancing markets should average to zero over time. Moreover, ISOs have proposed mechanisms to ensure imbalances average to zero over specific time intervals to enable participation of energy constrained resources like storage \cite{REMdraft, beaconenergy2011,NYISO2011}.

We consider a hypothetical balancing market in which each storage first commits a capacity, $S_i$, and then sets a price, $p_i$, which, depending on the market format, can be for energy or capacity. Our model assumes firms are paid the price they bid rather than all firms seeing the same price, which is also referred to as a discriminatory price auction. The merits of discriminatory versus uniform price auctions in power markets have been analyzed extensively, cf. \cite{Kahn2001dil,kremer2004role,Fabra2006auctions}, but no consensus on which format is superior has emerged.

In each time period, the system wide energy imbalance $B^t$, which we assume to be random, is allocated price-wise among the storages up to the their capacities. When $p_i$ is an energy price, it persists over a sequence of time periods. This represents the realistic scenario that imbalances are realized and physically responded to faster than the duration between markets, e.g., every few seconds; we describe this in detail in the next section. This setup is similar to the California Independent System Operator's (CAISO) real-time energy market \cite{caiso_bpm_2014}, as discussed in Section~\ref{capvsen}. Note that even if the energy imbalances exceed the combined storage capacity, there is virtually always sufficient generator reserves, which here we model simply as a more expensive storage with infinite capacity.

We remark that setting capacities prior to prices is an important difference between our setup and those used by some ISOs, wherein capacities and prices are bid simultaneously. Some theoretical results are given by \cite{Acemoglu2009price}, which finds that simultaneous bidding under deterministic demand never leads to pure strategy Nash equilibria. Further discussion on two-dimensional bidding in electricity markets is given by \cite{oren1994bidder,wilson2002multi}. We justify our modeling both as more tractable approximation to the case of simultaneous bidding, and as a potentially realistic model in its own right. In particular, system operators often require capacity information earlier than price information, as is the case with CAISO's residual unit commitment \cite{caiso_bpm_2014}. In a balancing market with storage, for instance, a system operator may require a capacity commitment every few hours, but allow firms to update their prices to reflect current conditions on shorter intervals.

\section{Price and capacity competition}\label{s:cthenp}
In this section, we develop tools for analyzing balancing markets with storage. First, we generalize Bertrand-Edgeworth price competition so that a single price bid determines the demand allocations for multiple dynamically coupled periods. We completely characterize the Nash equilibrium of this game by building on theoretical techniques from \cite{Acemoglu2009price}. Subsequently, we obtain new results for Nash equilibria when firms strategically set capacities prior to the price-setting stage when the demand is random. We primarily focus on duopolies for analytical tractability. All proofs are collected in the Appendix.

Throughout this section, we refer to the quantity $B$ as the sequence of imbalances. We examine special cases of the general framework in Section~\ref{AEM}, where we identify $B$ either as a sequence of energy imbalances to be allocated amongst the storages or as a demand for (deterministic) storage capacity.

\subsection{Price-setting stage}\label{dynprice}
We define our model for price-setting and how the resulting prices determine the allocation of random imbalances among the firms in Section~\ref{BENGR}. In Section~\ref{sec:priceeq}, we obtain the explicit mixed-strategy Nash equilibrium for this game. We assume here that the firms have fixed capacities $S_i$, $i=1,...,N$. Later in Section~\ref{capcomp}, the capacities themselves become the firms' strategic decisions, and the resulting payoff depends on the price equilibrium obtained in this section.

\subsubsection{Imbalance allocation and payoff rules.}\label{BENGR}
Each firm sets a price, $p_i\geq0$, which persists through all time periods, $t=0,...,T$. We refer to $T$ as the \textit{market horizon}. Between periods, firm $i$ leaks a constant fraction of its state of charge, $1-\alpha_i\in[0,1]$. In each period $t$, the system operator observes the random imbalances $B^t\in\mathbb{R}$, but is not aware of future demands. We assume that the sequence of imbalances are statistically described by the joint distribution $f(B)$. The case when $B$ is not random is simply when $f(B)$ is a discrete distribution with its entire mass on one point. 

In each period $t$, the imbalance $B^t\in\mathbb{R}$ is allocated amongst the firms according to the below linear program, which is parametrized by the price vector $p$.
\begin{eqnarray}
\underset{X_1^t,...,X_N^t}{\textrm{minimize}} && \sum_{i=1}^{N+1} p_i\left|X_i^t\right| \label{allo1}\\
\textrm{such that}&& \sum_{i=1}^{N+1} X_i^t = B^t\label{allo2}\\
&& 0 \leq \alpha_i S_i^t(p)+X_i^t\leq S_i,\quad i=1,...,N \label{allo3}
\end{eqnarray}
Here, the storages are dispatched in \textit{merit order}, i.e., the storage with the lowest price is used up to its capacity, then that with the second lowest, and so on until $B^t$ has been completely allocated in each time period. $X_i^t$ is the portion of $B^t$ allocated to firm $i$, and $S_i^{t}(p)$ its state of charge in period $t$. We display the dependence of the allocation on the price vector $p$ by denoting $\mathcal{X}_i^t(p)=X_i^t$.

Observe that (\ref{allo1})-(\ref{allo3}) allocates the imbalance in a given time period with no regard to future imbalances. This assumption is appropriate for our current scope because future imbalances are random and hence unknown to the system operator. Note that the system operator could achieve better performance by utilizing the statistics of future imbalances, i.e., $f(B)$, in its allocation decisions; unfortunately, this would make our model intractable. On the other hand, we could tractably model the simultaneous allocation of the entire sequence $B^1,...,B^T$, but this would unrealistically assume that the system operator can perfectly predict all future imbalances.

Firm $i$'s state of charge evolves according to
\begin{equation}
S_i^{t+1}(p)=\alpha_iS_i^{t}(p) + \mathcal{X}_i^t(p),\quad S_i^0(p)=S_i^0.\label{Sevo}
\end{equation}
Firm $i$'s total realized profit after the pricing game is then 
\[
p_i\sum_{t=0}^T\left|\mathcal{X}_i^t(p)\right|.
\]
Note that this payment implies that firms are paid identically for positive and negative imbalances. As discussed in Section~\ref{storageback1}, this is a reasonable assumption because both positive and negative imbalances are detrimental to system operation, e.g., both can lead to frequency instability.

Firm $N+1$ corresponds to conventional generator reserves, which cost $p_{N+1}=R$. We assume the capacity of conventional reserves is much larger than the storage pool so that $S_{N+1}=\infty$. This implies that $p_i\leq R$ if firm $i$ is to receive any payment; for this reason, $R$ is sometimes referred to as the \textit{reservation utility}. Under price ties, firms are randomly assigned priority, e.g., if $p_i=p_j$ for some $i\neq j$, firm $i$ receives first allocation with probability one half. A variety of tie-rules could be specified that would serve equivalently in our analysis, and the particular choice does not affect our results.

Let $\mu_i$ be firm $i$'s mixed strategy, which we assume to be independent of the other firms' strategies. Let $\mathbb{E}$ denote the expectation over $B$ throughout the paper. The payoff associated with $\mu_i$ is
\begin{eqnarray}
\pi_i(\mu)&=&\int_{p}\left(\prod_{j}\mu_j(p_j)\right)p_i\mathbb{E}\left[\sum_{t=0}^T\left|\mathcal{X}_i^t(p)\right|\right]dp.\label{rho1}
\end{eqnarray}
$\mu$ is a mixed strategy Nash equilibrium if for each firm $i$, $\pi_i(\mu)\geq \pi_i(\mu_i',\mu_{-i})$ for all $\mu_i'$. We denote the payoff of a pure strategy $\pi_i(p_i,\mu_{-i})$. Note that this definition is essentially the Bayesian equilibrium where $B$ is the only unknown parameter and all firms share the same knowledge \cite{fudenberg1991GT,osborne1994course}.

\subsubsection{Price equilibrium.}\label{sec:priceeq}
In this section, we characterize the price equilibrium. Discontinuous games can fail to have a (pure or mixed) Nash equilibrium \cite{maskin1986theory}. The next lemma guarantees that the pricing game always has a mixed-strategy price equilibrium.
\begin{lemma}\label{MSEexist}
A mixed-strategy price equilibrium always exists.
\end{lemma}
The below lemma gives the explicit, pure strategy equilibrium in two extreme cases.
\begin{lemma}\label{pricemix}
In the following two cases, the mixed strategy equilibrium reduces to a pure strategy equilibrium.
\begin{enumerate}
\item If for each $i$, $\mathbb{E}\left[\sum_{t=0}^T\left|\mathcal{X}_i^t(p)\right|\right]=0$ 
when $i$'s price is maximal ($i=\underset{j}{\textrm{argmax}}\;p_j$), the pure strategy equilibrium is $p_i=0$ for all $i$.
\item If $\mathbb{E}\left[\sum_{t=0}^T\left|\mathcal{X}_i^t(p)\right|\right]=\mathbb{E}\left[\sum_{t=0}^T\left|\mathcal{X}_i^t(p')\right|\right]>0$ for all $i$ and all price vectors $p$ and $p'$, the pure strategy equilibrium is $p_i=R$ for all $i$.
\end{enumerate}
\end{lemma}
In standard Bertrand-Edgeworth competition with one period and deterministic demand, the first case reduces to the firms undercutting each other's price to zero, and the second case to the firms setting the maximum price because they are guaranteed that their capacity will be exhausted. When the imbalance, $B$, is random, the first case entails that any realization of $B$ be containable by any size $N-1$ subset of the firms, and the latter that any realization exhausts each firm's capacity in every time stage.

The following lemma characterizes the mixed strategies in the general $n$-firm case.
\begin{lemma}\label{Nfirms}
Let $[L_i,U_i]$ denote the support of firm $i$'s mixed strategy, $\mu_i(p)$.
\begin{enumerate}
\item For any firm $j$'s support, $[L_j,U_j]$, there are at least two firms $i$ for which $\mu_i(p)>0$ for all $p\in [L_j,U_j]$.
\item Only one atom can exist across all firms, and if one does, it must be at $U=\max_iU_i$.
\item $U=R$.
\end{enumerate}
\end{lemma}

We henceforth restrict our attention the more tractable two-firm case. Without loss of generality, we assume that $S_1\geq S_2$. We use the notation $[X]_Y^Z$ to denote the quantity $X$ truncated below by $Y$ and above by $Z$, i.e., $[X]_Y^Z=\min\{\max\{X,Y\},Z\}$. Let $\underline{\mathcal{X}}_i$ be the sum of the magnitudes of the imbalances captured by firm $i$ when it has first priority in the allocation of $B^t$ in each period, and $\overline{\mathcal{X}}_i$ when it has second priority. Mathematically, let $p'$ denote a price vector in which $p_i<p_{-i}$, and $p''$ vice versa. Then
\begin{eqnarray*}
\underline{\mathcal{X}}_i&=&\mathbb{E}\left[\sum_{t=0}^T\left|\mathcal{X}_i^t\left(p'\right)\right|\right] \\
&=& \mathbb{E}\left[\sum_{t=0}^T\left| \left[B^t\right]_{-S_i^t\left(p'\right)}^{S_i-S_i^t\left(p'\right)} \right|\right]
\end{eqnarray*}
and
\begin{eqnarray*}
\overline{\mathcal{X}}_i&=&\mathbb{E}\left[\sum_{t=0}^T\left|\mathcal{X}_i^t\left(p''\right)\right|\right]\\
&=& \mathbb{E}\left[\sum_{t=0}^T\left| \left[B^t-\left[B^t\right]_{-S_{-i}^t\left(p''\right)}^{S_{-i}-S_{-i}^t\left(p''\right)}\right]_{-S_i^t\left(p''\right)}^{S_i-S_i^t\left(p''\right)} \right|\right].
\end{eqnarray*}
It is straightforward to show that $\underline{\mathcal{X}}_i\geq\overline{\mathcal{X}}_i$, $\underline{\mathcal{X}}_1\geq\underline{\mathcal{X}}_2$, and $\overline{\mathcal{X}}_1\geq\overline{\mathcal{X}}_2$.

Note that $\underline{\mathcal{X}}_i$ and $\overline{\mathcal{X}}_i$ can be directly computed from each $S_i$, $\alpha_i$ and the distribution of $B$. For example, when there is only one time period and $S_i^0=0$, we have
\begin{eqnarray}
\underline{\mathcal{X}}_i &=& \int_{0}^{S_i} B f(B)dB + S_i(1-F\left(S_i\right))\label{Brandlow}\\
\overline{\mathcal{X}}_i &=& \int_{0}^{S_i} B f\left(B+S_{-i}\right)dB + S_i\left(1-F\left(S_1+S_2\right)\right)\label{Brandhigh}
\end{eqnarray}
When $B$ is also not random,
\begin{eqnarray}
\underline{\mathcal{X}}_i &=& \left[B\right]_{0}^{S_i}\label{Bdetlow}\\
\overline{\mathcal{X}}_i &=& \left[B-\left[B\right]_{0}^{S_{-i}}\right]_{0}^{S_i}\label{Bdethigh}
\end{eqnarray}

The following new result explicitly characterizes the pricing equilibrium for general duopolies.
\begin{theorem}\label{prop:mu}
Assume without loss of generality that $S_1\geq S_2$. The equilibrium payoffs are given by
\begin{eqnarray*}
\pi_1^*&=& R\overline{\mathcal{X}}_1\\
\pi_2^*&=& \frac{R\overline{\mathcal{X}}_1\underline{\mathcal{X}}_2}{\underline{\mathcal{X}}_1}
\end{eqnarray*}
The mixed strategies are
\begin{eqnarray*}
\mu_1(x)&=&\frac{R\overline{\mathcal{X}}_1\underline{\mathcal{X}}_2}{\underline{\mathcal{X}}_1\left(\underline{\mathcal{X}}_2-\overline{\mathcal{X}}_2\right)x^2}+\frac{\underline{\mathcal{X}}_2\overline{\mathcal{X}}_1-\overline{\mathcal{X}}_2\underline{\mathcal{X}}_1}{\underline{\mathcal{X}}_1\left(\underline{\mathcal{X}}_2-\overline{\mathcal{X}}_2\right)}\delta(x-R),\quad x\in [L,R]\\
\mu_2(x)&=&\frac{R\overline{\mathcal{X}}_1}{\left(\underline{\mathcal{X}}_1-\overline{\mathcal{X}}_1\right)x^2}, \quad x\in[L,R]
\end{eqnarray*}
where $\delta(x)$ is the Dirac delta function and $L=R\overline{\mathcal{X}}_1/\underline{\mathcal{X}}_1$.
\end{theorem}

Observe that the equilibrium in Theorem~\ref{prop:mu} is unique because our derivation constructively identifies the properties an equilibrium must have. It is straightforward to show that the for two firms Lemma~\ref{pricemix} is a special case of Theorem~\ref{prop:mu} by considering two limiting cases. In the first, $\overline{\mathcal{X}}_i=0$, $\pi_1^*=\pi_2^*=0$, and each $\mu_i(x)$ has an atom of mass one at $x=0$. In the latter, $\overline{\mathcal{X}}_i=\underline{\mathcal{X}}_i$, $\pi_1^*=R\overline{\mathcal{X}}_1$, $\pi_2^*=R\overline{\mathcal{X}}_2$, and each mixed strategy has an atom of mass one at $x=R$.

\subsection{Capacity-setting stage}\label{capcomp}
We now turn to the strategic determination of storage capacities; recall that this decision stage occurs prior to price-setting. For analytical tractability, we restrict our attention to a single time period, and consider both the cases when $B$ is deterministic and random. Since there is no dynamic coupling over a single time period, we can assume $B\geq 0$ and $S_i^0=0$ for both firms without loss of generality. In these cases, the price equilibria are given by (\ref{Brandlow})-(\ref{Bdethigh}) and Theorem~\ref{prop:mu}. For tractability, we only seek pure strategy capacity equilibria, as is standard in almost all supply function-based approaches, cf. \cite{Bolle1992supply,green1992british,hobbs2002supply,baldick2004lin}.

\subsubsection{Capacity game}
Prior to price setting, the firms choose capacities $S_1, S_2\geq0$ to maximize expected profits. Each firm incurs an opportunity cost as discussed in Section~\ref{storageback}, $-\gamma_iS_i$, which must satisfy $0<\gamma_i \leq R$ for the game to be nontrivial. The price equilibrium payoff of the larger firm is denoted by $\overline{\pi}(S_i,S_{-i})$ and of the smaller firm by $\underline{\pi}(S_i,S_{-i})$, where $S_i\geq S_{-i}$, i.e., the first argument is the larger capacity. The expected net profit of firm $i$ is the payoff from the pricing stage minus the opportunity cost:
\begin{eqnarray*}
\psi_i(S)&=& -\gamma_iS_i+\left\{ \begin{array}{ll}\overline{\pi}\left(S_i,S_{-i}\right) & \indent \textrm{if }S_i\geq S_{-i}\\
\underline{\pi}\left(S_{-i},S_{i}\right) &\indent \textrm{if }S_i< S_{-i}
 \end{array}\right.
\end{eqnarray*}
The capacity game is defined by both firms simultaneously maximizing their respective payoffs, $\psi_1(S)$ and $\psi_2(S)$. A capacity pair $S$ is a Nash equilibrium if it satisfies the standard definition:
\begin{equation}
\psi_i\left(S\right)\geq\psi_i\left(S_i',S_{-i}\right)\quad \forall\; S_i'\geq0,\; i=1,2.\label{capeqdef}
\end{equation}

\subsubsection{Capacity equilibrium}
The below result for the case that $B$ is deterministic and its $n$-firm generalization are given in \cite{Acemoglu2009price}.
\begin{lemma}\label{M1Cap}
When $B$ is deterministic, the set of capacity equilibria is given by the continuous range of values described by
\begin{displaymath}
S_1+S_2=B,\quad\frac{R-\gamma_{i}}{2R-\gamma_i}B\leq S_{i} \leq S_{-i},\quad i=1,2.\label{capeqM1}
\end{displaymath}
\end{lemma}
Observe here that the sum of the firms' capacities exactly equals the total demand, $B$, which, according to the first part of Lemma~\ref{pricemix}, means that at equilibrium both firms will set their price at the maximum, $R$.

We now give a novel characterization of the capacity equilibrium when $B$ is random and nonnegative.
\begin{lemma}\label{capexist}
Suppose that that $B$ is random and nonnegative and that $f(B)$ is positive and decreasing for $B\geq0$. Then $\overline{\pi}(S_i,S_{-i})-\gamma_iS_i$ and $\underline{\pi}(S_{-i},S_i)-\gamma_{i}S_{i}$ are respectively strictly concave and strictly quasiconcave in $S_{i}$.
\end{lemma}
Because the segments of $\psi_i(S)$ where $S_i< S_{-i}$ and $S_i>S_{-i}$ are both quasiconcave and differentiable in $S_i$, pure strategy capacity equilibria are liable to occur where their derivatives are zero, enabling us to obtain constructive characterizations. Note, however, that $\psi_i(S)$ may not be quasiconcave due to the kink where the two segments meet at $S_i=S_{-i}$, and in some cases a pure strategy capacity equilibrium may not exist.

Let the cumulative distribution of $B$ be given by $F(B)=\int_{0}^Bf(x)dx$. Denote the inverse cumulative distribution $F^{-1}(x)$, let $F^{-1}(0)=0$, and note that $F^{-1}(x)$ is well-defined for $x\in[0,1)$ because $f(B)>0$ for $B\geq0$. Define
\begin{eqnarray}
\Lambda^1_i&=&F^{-1}(1-\gamma_i/R)\\
\Lambda^2_{i}&=&S_i  \textrm{ such that } \left.\frac{d\underline{\pi}\left(S_{-i},S_{i}\right)}{dS_{i}}\right|_{S_{-i}=\Lambda^1_{-i}-S_{i}}=\gamma_{i}\label{Sismalleq}
\end{eqnarray}
If $S_i\geq S_{-i}$, setting the derivative of $\psi_i\left(S\right)=\overline{\pi}(S)-\gamma_iS_i$ with respect to $S_i$ equal to zero gives $S_i+S_{-i}=\Lambda^1_i$. If $S_i<S_{-i}$, setting the derivative of $\psi_i\left(S\right)=\overline{\pi}(S)-\gamma_iS_i$ to zero gives $S_{i}= \Lambda^2_{i}$.

Define
\begin{eqnarray*}
\hat{S}^i&=&\left(\Lambda_i^1-\Lambda^2_{-i},\Lambda^2_{-i}\right).
\end{eqnarray*}
$\hat{S}^i$ is a pure strategy Nash equilibrium if it satisfies the definition given by (\ref{capeqdef}). Note that we have not proven that $\Lambda^2_{i}$ is uniquely defined, i.e., that there is a single solution to (\ref{Sismalleq}). However, this is the case in all of our numerical examples in Section~\ref{capvsen}.

\begin{lemma}\label{M2capeqs}
Suppose that that $B$ is random and nonnegative and that $f(B)$ is positive and decreasing. Then the $\hat{S}^i$, $i=1,2$ are the only possible pure-strategy equilibria.
\end{lemma}
We base our analysis in Section~\ref{capvsen} on Lemma~\ref{M2capeqs} by only considering the equilibria $\hat{S}^i$, $i=1,2$. We remark that since $\psi_i(S)$ is continuous, a mixed-strategy equilibrium must exist even if both of the $\hat{S}^{i}$ fail to be equilibria \cite{Glicksberg1952Nash}. In our subsequent analysis in Section~\ref{capvsen}, we limit our attention to the above pure strategy equilibria, at least one of which always exists in our examples in Section~\ref{capvsen}.
\section{Analysis of electricity markets}\label{AEM}
In this section, we apply our results on price and capacity competition to energy imbalance markets with storage. We first restrict our attention to markets with energy-based pricing and examine the effect of market horizon and leakage. We then use capacity competition to compare markets with capacity and energy-based payments. As discussed in Section~\ref{storageback}, prices and profits beyond that necessary to recoup opportunity costs in our model are the exclusive result of strategic behavior. Hence, in this section we will equate better performance with lower firm prices and profits.

Our application of price and capacity competition most closely parallels supply function modeling of generator competition (see the introduction) in that here firms submit what is essentially a two-parameter bid curve consisting of a single price and a capacity limit. As discussed in Section~\ref{storageback}, the price and capacity competition framework is highly appropriate for storage because it precisely captures its hard capacity limits, dynamic coupling, and low operating costs while admitting more nuanced analyses than would be obtainable from generally supply function competition.

\subsection{Dynamic effects}
In this section, we use the results of Section~\ref{dynprice} to examine the dynamic effects of the market horizon and physical efficiency (leakage) under energy-based pricing. Here we assume that capacities are given, i.e., not set strategically in a prior stage. Note that although our formulation can accommodate serial correlations in the sequence of energy imbalances, we do not consider their effects for concision and because we regard market horizon and efficiency to be more influential parameters.

\subsubsection{Market horizon}
We examine the effect of the market horizon, $T$, as defined in Section~\ref{BENGR} on equilibria. Note that this relationship can only be observed through energy pricing because the number of periods does not (directly) affect firm profits under capacity payments. We look at the expected equilibrium prices, which can be calculated analytically from Theorem~\ref{prop:mu} to be
\begin{eqnarray*}
\rho_1^*&=&\frac{R}{\underline{\mathcal{X}}_1\left(\underline{\mathcal{X}}_2-\overline{\mathcal{X}}_2\right)}\left(\overline{\mathcal{X}}_1\underline{\mathcal{X}}_2\left(1+\ln\left(\frac{\underline{\mathcal{X}}_1}{\overline{\mathcal{X}}_1}\right)\right)-\overline{\mathcal{X}}_2\underline{\mathcal{X}}_1\right)\\
\rho_2^*&=&\frac{R\overline{\mathcal{X}}_1}{\underline{\mathcal{X}}_1-\overline{\mathcal{X}}_1}\ln\left(\frac{\underline{\mathcal{X}}_1}{\overline{\mathcal{X}}_1}\right).
\end{eqnarray*}

\begin{figure}[h]
\centering
\includegraphics[scale=1]{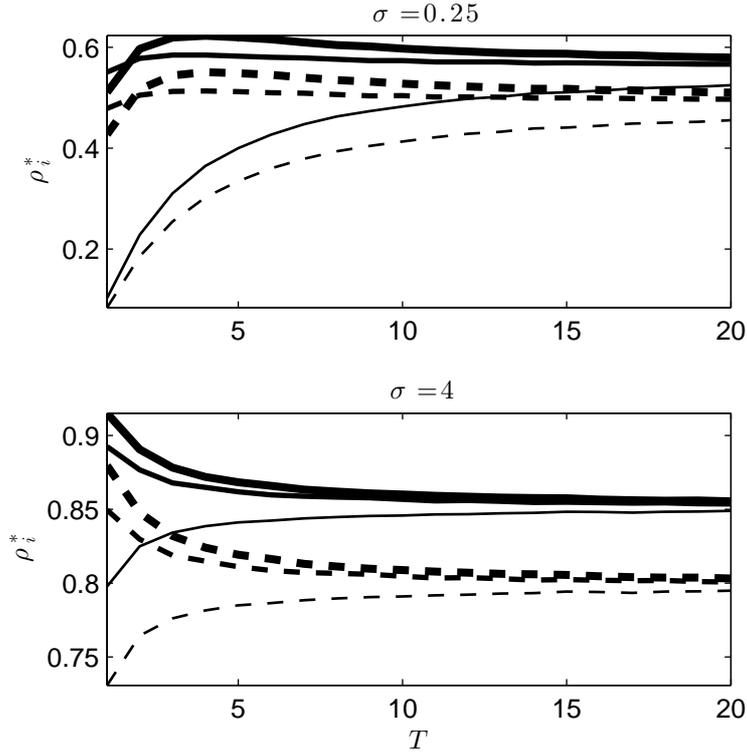} 
\caption{Expected equilibrium prices as a function of horizon, $T$, with the large firm solid and the small firm dashed. The energy imbalance in each period has variance $\sigma=1/4$ in the top plot and $\sigma=4$ in the bottom. Each plot depicts the cases when the initial states of charge are zero, one quarter, and half their capacities ($S^0=\{0,S/4,S/2\}$), with thicker lines indicating higher initial states of charge.}
\label{fig:DynPriceEq}
\end{figure}

Figure~\ref{fig:DynPriceEq} shows $\rho_1^*$ and $\rho_2^*$ as a function of the market horizon, $T$. The firm parameters are $S_1=1.5$ and $S_2=1$ and $\alpha_1=\alpha_2=1$. Each period's energy imbalance is drawn from an independent, zero mean Gaussian distribution with variance $\sigma$, where $\sigma=1/4$ in the top plot and $\sigma=4$ in the bottom. Each plot shows the result for three different initial states of charge, which we assume satisfy $S_1^0/S_1=S_2^0/S_2$. This assumption is consistent with mechanisms used by system operators to `reset' storage to ensure that imbalances average to zero over time \cite{REMdraft, beaconenergy2011,NYISO2011}. The integrals necessary to compute the equilibria were evaluated numerically using $10^5$ Monte Carlo points.

As is to be expected, regardless of initial condition the expected prices converge to the same long-horizon value for each storage, which increases with $\sigma$. However, the initial condition can lead to qualitatively different transient behaviors. When the firms begin empty ($S^0=0$), each $\rho_i^*$ monotonically increases with $T$. This is because the firms cannot capture profits when the early imbalances are negative, and thus have a relatively high probability of not exhausting their capacity, leading to undercutting. In numerical experiments, we observed that this behavior only occurs when the initial state of charge is under a tenth the total capacity (or within a tenth of full), and thus does not account for the majority of cases.

When the initial state of charge is between ten and ninety percent of the capacity, we observe three different equilibrium regimes as $T$ is increased:
\begin{enumerate}
\item \textit{(Small $T$)} In this range, each firm rarely hits its energy capacity limit before $T$. Consequently, either firm can physically accommodate most of the demand, causing them to engage in undercutting. This equilibrium is closer to the first case in Lemma~\ref{pricemix}.
\item \textit{(Medium $T$)} As $T$ is increased, firms are likely to have their capacities exhausted exactly once regardless of their price. Consequently, firms set higher prices closer to the reservation utility. This equilibrium is closer to the second case in Lemma~\ref{pricemix}.
\item \textit{(Large $T$)} Further increasing $T$ causes the firms to enter a regime where the lower priced firm cycles faster, i.e., hits its capacity limits more than the other, and hence profits more. As a result, the firms resume undercutting. There is a `knee' in the curve, beyond which increasing $T$ further contributes little to economic competitiveness.
\end{enumerate}

In the top plot, the first regime appears within the first couple periods, the second between three and seven periods, and the third past the seventh period. In the bottom plot, only the latter two regimes are present, with the transition occurring around the third period. The knee at which a larger market horizon does not significantly reduce firm profits occurs around $T=20$ in the top plot and $T=7$ in the bottom. Generally, the location of the knee can be expected to increase as a function of storage size and decrease as a function of energy imbalance variance.

Market horizon is an important design parameter for balancing markets because too large a horizon will make adapting to changes difficult and too short a horizon could add price volatility. For the majority of initial conditions, the expected prices decrease with $T$. Therefore, the market horizon should be chosen to be large enough that the expected prices have nearly converged to the long-horizon value. From Figure~\ref{fig:DynPriceEq}, we see that this convergence occurs relatively early at the knee in each plot. Hence, our analysis enables the identification of market horizons that are small but still minimize gaming.

\subsubsection{Leakage}
We now consider the effect of losses due to the leakage parameter, $\alpha$, which was defined in Section~\ref{storageback}. Like the market horizon, leakage can only be analyzed in a dynamic setting because it manifests between periods. As discussed in Section~\ref{storageback}, we use leakage as an analytically tractable proxy for general inefficiencies like injection and extraction losses, though we note there are important differences between the two mechanisms; we leave detailed analysis of the effects of inefficiency to future work.

\begin{figure}[h]
\centering
\includegraphics[scale=1]{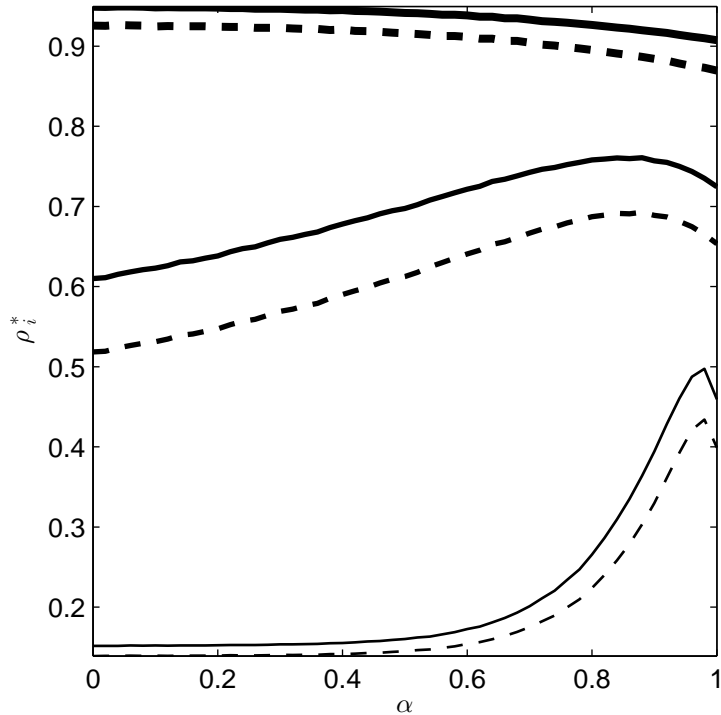} 
\caption{Expected equilibrium price as a function of leakage. The three pairs of curves correspond to the variances $\sigma=\{1/10,1,10\}$, with thicker lines corresponding to larger $\sigma$.}
\label{fig:DynPriceEqLeak}
\end{figure}

We again study the expected equilibrium price, $\rho_i^*$, for $T=100$, which represents the long-horizon value. The firms have capacities $S_1=1.5$ and $S_2=1$ with $S_i^0=S_i/2$. Figure~\ref{fig:DynPriceEqLeak} shows $\rho_i^*$ as a function of $\alpha$ for $\sigma=\{1/10,1,10\}$, now with thicker lines corresponding to larger variances.

At the left side of the plot, the state of charge drops to zero at the end of each time period, effectively resulting in a single period model. Consequently, at low variances, the firms undercut intensely, and at high variances where capacity is expected to be exhausted, the firms set prices near the reservation utility. We remark that inefficient storage is less useful to power systems due to high energy waste, which is not penalized in our model. As the physical efficiency improves, the prices increase in the lower variance cases, and then decrease just before reaching zero leakage. This transition from increasing to decreasing prices occurs because the profit gain from cycling more frequently by being the lower priced firm begins to outweigh the profits lost to leakage. On the other hand, in the high variance case, the price decreases monotonically as the profit gains from cycling more frequently overtake those of charging high prices in each individual period.

From a physical perspective, losses are always undesirable, and modern batteries can achieve roundtrip efficiencies of over 70\% \cite{Dunn2011choices} and up to 95\% for technologies like superconducting magnetic energy storage \cite{Ibrahim2008char,Castillo2014survey}. Therefore, focusing on the right side of the plot, we conclude that lower leakage reduces gaming. More precisely, lower leakage strengthens the state of charge's dynamic coupling. This increases the value of being cycled more frequently relative to setting a high price in each individual period, thus inducing the firms to lower their prices. This conforms with the results of \cite{taylor2011store}, which also finds through an inventory control-based model that increased leakage reduces competition. Note that this is a similar outcome to that achieved by increasing the market horizon.

\subsection{Pricing capacity versus energy}\label{capvsen}
Under capacity pricing, firms are paid for committed capacity in forward markets (FM). Under energy pricing, firms are paid for absorbing energy imbalances in real-time markets (RTM). The key mathematical difference between each format is that the system operator's capacity requirement is deterministic and energy imbalances are random. In this section, we use the capacity competition results from Section~\ref{capcomp} to compare these two formats. Our analysis here is highly idealized because the results in Section~\ref{capcomp} assume one stage of competition, meaning that storages are competing to absorb a single energy imbalance.

Tables~\ref{tab:M1} and \ref{tab:M2} summarize the sequence of events under each type of pricing. In both cases, firms first set their maximum capacities, $S_i$, in the FM. Under capacity pricing, firms then set prices for deterministic capacity (also known as `capacity premiums') in the FM, and the system operator purchases a portion of each firm's capacity at these prices in the FM. The equilibrium in this case is described by Lemma~\ref{M1Cap}. Under energy pricing, firms set prices for energy (also known as `imbalance fees') in the FM, and firms are paid for absorbing energy imbalances in the RTM. The equilibrium in this case is described by Lemma~\ref{M2capeqs}. While some energy markets allow prices and capacities to be set in the RTM, our assumption is consistent with markets like CAISO's real-time (5-minute) energy market, in which capacities and prices are set on the hour, and generators are dispatched price-wise every 5-minutes \cite{caiso_bpm_2014}.

\begin{table}[htb]
\caption{Sequence of events under capacity pricing. The capacity equilibrium in this case is described by Lemma~\ref{M1Cap}.}
		\label{tab:M1}
	\centering
			\begin{tabular}{c|c|l|c}
			\hline
			Stage 	&  Time & Description & Competition \\
			\hline
			1		& FM 	& Firms set maximum capacities, $S_i$. & Capacity \\

			2	 	& FM 	& Firms set capacity prices, $p_i$. & Price \\

			3		& FM	& Deterministic capacity requirement, $B$, allocated price-wise. & --  \\
\hline
		\end{tabular}
\end{table}

\begin{table}[htb]
\caption{Sequence of events under energy pricing. The capacity equilibrium in this case is described by Lemma~\ref{M2capeqs}.}
		\label{tab:M2}
	\centering
			\begin{tabular}{c|c|l|c}
			\hline
			Stage 	&  Time 	& Description & Competition \\
			\hline
			1 		& FM  	& Firms set maximum capacities, $S_i$. & Capacity \\

			2 		& FM 	& Firms set energy prices, $p_i$. & Price \\ 

			3 		& RTM	& Random energy imbalance, $B$, allocated price-wise. & -- \\
\hline
		\end{tabular}
\end{table}

We first compare the qualitative properties of the two formats described in Tables~\ref{tab:M1} and \ref{tab:M2}. Ascribing Lemma~\ref{M1Cap} to markets with capacity pricing, firms will declare capacities that sum exactly to the total capacity requirement. Consequently, the price equilibrium reduces to the second scenario in Lemma~\ref{pricemix}, which says that each firm's pure strategy is to bid the maximum price, $p_1=p_2=R$, the least economically competitive outcome. Moreover, there is a potentially wide range of capacity equilibria, some of which may be inequitable to some firms \cite{Acemoglu2009price}.

Energy-based markets are described by Lemma~\ref{M2capeqs} on capacity equilibria when the demand in the pricing stage is random. Under the conditions of the lemma, at most two pure-strategy capacity equilibria can exist, but in some scenarios there may only be mixed equilibria. This implies that firm capacity commitments are more predictable under energy pricing, and is reminiscent of a central result of \cite{klemperer1989supply} showing that uncertainty reduces the number of pure-strategy supply function equilibria. It is easy to see from Theorem~\ref{prop:mu} that the price equilibrium is mixed and result in expected firm profits below pricing at the maximum.

Next, we numerically compare the capacity equilibria under capacity and energy pricing. We assume that the real-time energy imbalance obeys the half-normal distribution $f(B)=\sqrt{2/\pi}e^{-B^2/2}$, $B\geq0$, because normal distributions have long been used to model mismatches in supply and demand in power systems \cite{billinton1996power}. The system operator's capacity requirement is just the mean of $B$, $\sqrt{2/\pi}$; we comment more on this choice later. In both market formats, the reservation utility, $R$, is based on the cost of generator reserves. Specifically, suppose that $x$ units of generator capacity costs $r(x)=ax^2+bx$. The reservation price for capacity is then $R_1=r(x)/x$, and for expected reserve energy $R_2=r(x)/\mathbb{E}[\min\{B,x\}]$. If we assume that all capacity will be procured from storage, i.e., $x=0$, we have
\begin{eqnarray*}
R_2&=&\lim_{x\rightarrow 0}\frac{r(x)}{E[\min\{B,x\}]}\\
&=&\lim_{x\rightarrow 0}\frac{2ax+b}{1-F(x)}\quad \textrm{by l'H\^{o}pital's rule}\\
&=&\lim_{x\rightarrow0}\frac{r(x)}{x}\\
&=&R_1
\end{eqnarray*}
Hence, we may assume that both formats have the same reservation utility. For simplicity, we set $R_1=R_2=1$.

\begin{figure}[h]
\centering
\includegraphics[scale=1]{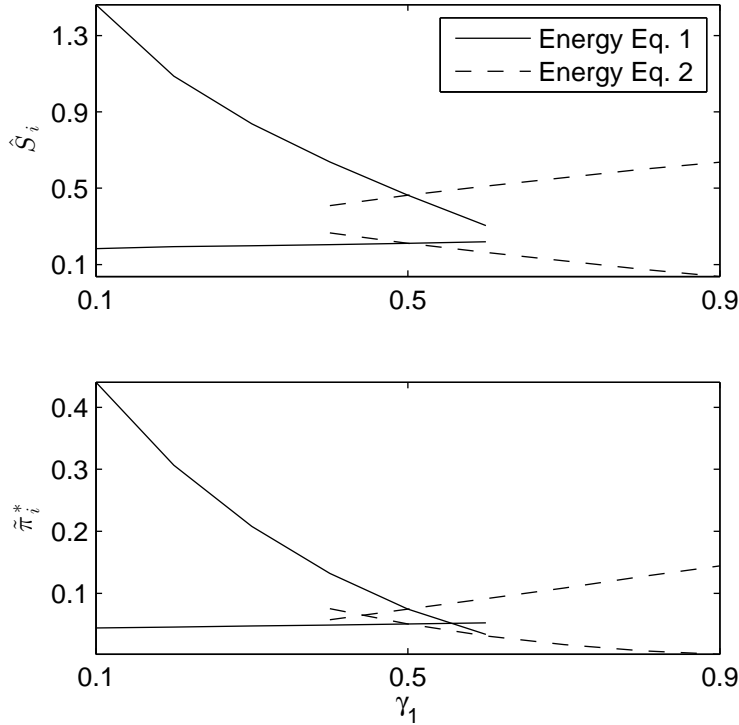} 
\caption{Each storage's equilibrium capacity and payoff for $\gamma_2=0.5$. The two equilibria are identical when $\gamma_1=\gamma_2=0.5$.}
\label{fig:S1M2CapGam}
\end{figure}
Figure~\ref{fig:S1M2CapGam} shows the pure strategy equilibria under energy pricing. Two equilibria can coexist, but only when the firms' opportunity costs are similar, i.e., $\gamma_1\approx\gamma_2$. Otherwise, the single equilibrium is $\hat{S}^i$ where $\gamma_i<\gamma_{-i}$, which corresponds to the firm with the lower opportunity cost committing a higher capacity. The associated equilibrium profits correspond to prices well below the reservation utility, $R=1$; for instance, when $\gamma_1\approx\gamma_2$, the profits are approximately one quarter the committed capacity, corresponding to average prices of approximately $1/4$. 

\begin{figure}[h]
\centering
\includegraphics[scale=1]{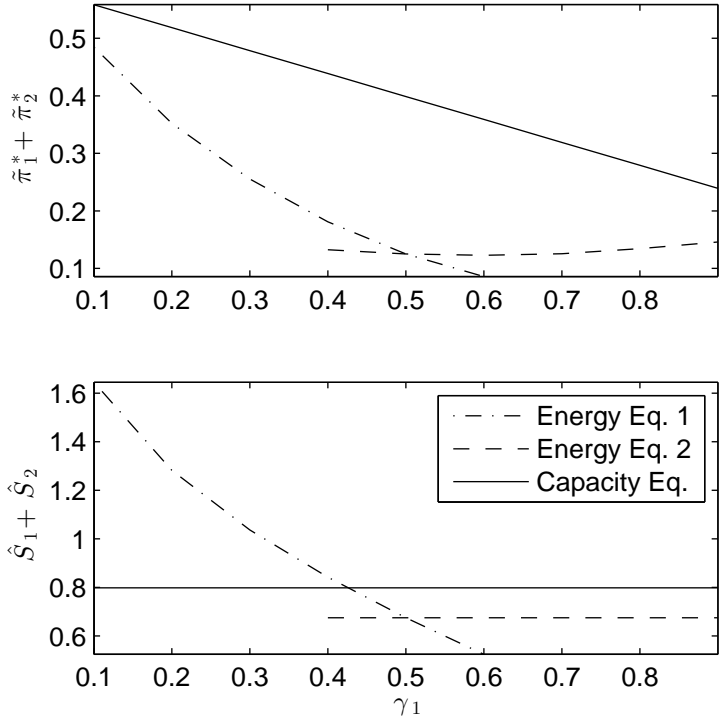} 
\caption{Total firm profits under energy and capacity pricing for $\gamma_2=0.5$.}
\label{fig:S1M1M2ProfGam}
\end{figure}

Figure~\ref{fig:S1M1M2ProfGam} shows the total profits and capacity commitment over the same range as Figure~\ref{fig:S1M2CapGam}, along with the deterministic capacity pricing case. Here, profits under capacity pricing are substantially higher, even in the low opportunity cost regime where energy pricing leads to greater capacity commitment. We remark that the system operator would typically procure far more than the mean, $\sqrt{2/\pi}$, so we regard the capacity commitment and profit under capacity pricing to be a conservative approximation.
%

The salient differences between the formats are summarized below:
\begin{itemize}
\item Capacity pricing leads to larger capacity commitments and consistent (pure strategy), high prices.
\item Energy pricing leads to smaller capacity commitments and inconsistent (mixed strategy), lower prices.
\end{itemize}
Except for the randomization of prices under energy pricing, these are well-known characteristics in other markets (e.g., conventional energy and reserves), and reflect the tradeoff between robust (capacity-based) approaches that guarantee adequate capacity and average (energy-based) cost approaches that achieve better economic competitiveness (see, e.g., \cite{oren2000cap} or \cite{Cramton2005cap}). To our knowledge, this is the first such analysis capturing this tradeoff in balancing markets with storage.

\section{Conclusion and future work}\label{conc}
Energy storage and aggregations of shiftable loads could soon compete alongside traditional generators in energy balancing markets. As in traditional generation markets, storage may have opportunities to exert market power, justifying analysis of strategic behavior in these new markets. Price and capacity competition is a game theoretic model of strategic behavior that precisely captures three essential characteristics of storage: hard capacity limits, dynamically coupled energy states, and low marginal costs. Using our new results on price and capacity competition, we obtain the following insights about designing energy balancing markets with storage.
\begin{itemize}
\item Energy pricing leads to lower, randomized prices and lower capacity commitments. Capacity pricing leads to higher, consistent prices and higher capacity commitments. These two formats represent opposing extremes of possible market designs, which may have both energy and capacity payments. Choosing how much to pay for energy versus capacity is therefore a tradeoff between economically efficient pricing with low capacity commitment and robust capacity commitment with high prices. Despite the fact that, as discussed in Section~\ref{storageback1}, high prices in our model are the result of strategic behavior, capacity pricing may nevertheless be useful when the objective is to encourage investment in storage and its participation in balancing markets. Likewise, energy pricing may be well-suited to maintaining the competitiveness of markets with established participants.
\item Longer market horizons and lower leakage (and, by proxy, higher physical efficiency) both reduce gaming by inducing firms to compete to be cycled more frequently by lowering their prices (under energy pricing). The benefits of a longer horizon are easily obtainable because equilibrium prices converge rapidly with horizon length.
\end{itemize}

We now discuss some relevant future research directions. First, we believe that there are many additional questions that could be addressed with the framework in this paper, for instance the effect of temporal correlations in the energy imbalances, and sequences of non-zero mean energy imbalances. Since we have assumed a discriminatory payment mechanism, it is of interest and would likely be a similar theoretical exercise to characterize a game based on a uniform-price payment mechanism. Barrier to entry is another important market feature; however, this would likely be more difficult to analyze because it entails a non-trivial extension of our framework to the $n$-firm case. Models in which price and capacity are bid simultaneously or in which hybrid payments are made for both price and capacity may be a better match for real markets, but may be harder to analyze because each firm's strategy space will be two-dimensional. Finally, a dynamic stochastic game framework like that of \cite{pakes1994nash} could enable the analysis of sequences of markets with price and capacity bidding.
\section*{Appendix}

\proof{(Lemma~\ref{MSEexist}).}
We show that a mixed strategy equilibrium exists in the pricing game. Existence for a single period when $B$ is not random is proved in Prop. 4.3 \cite{Acemoglu2007cong} using Theorem 5 of \cite{maskin1986theory}. We may straightforwardly adapt their approach to the present scenario. Let $\hat{\pi}(p)=\sum_{i=1}^N\pi_i(p)$. For each firm $i$, let $G_i\in \mathbb{N}$. For each $G$ with $0\leq G\leq G_i$ and $j\neq i$, $1\leq j\leq N$, let $g_{ij}^G$ be a one-to-one continuous function. Let $\hat{P}(i)$ be defined as
\begin{displaymath}
\hat{P}(i)=\left\{p\;|\; \exists\; j\neq i,\; \exists\; G,\; 0\leq G\leq G_i \textrm{ s.t. } p_j=g_{ij}^G(p_i)\right\}.
\end{displaymath}
Theorem 5 in \cite{maskin1986theory} is as follows:
\begin{theorem}
Suppose that $\pi_i(p_i,p_{-i})$ is bounded, continuous in $p$ except on a subset $P^*$ of $\hat{P}(i)$, and weakly lower semicontinuous in $p_i$ for all $i$, and that $\hat{\pi}(p)$ is upper semicontinuous in $p$. Then a mixed-strategy equilibrium exists.
\end{theorem}

Using the argument in \cite{Acemoglu2007cong,Acemoglu2009price}, it can be shown that $\pi_i(p)$ is bounded, continuous in $p$ except at a subset $P^*$, and weakly lower semicontinuous when $B$ is deterministic. All of these properties are preserved by taking the expectation over $B$ and thus hold for $\pi_i(p)$ when $B$ is random. Clearly, $\hat{\pi}(p)$ is continuous as well since it is the sum of continuous functions. This establishes the existence of a mixed-strategy equilibrium.
\endproof


\proof{(Lemma~\ref{pricemix}).}
We proceed case-wise as in the lemma.
\begin{enumerate}
\item Let $i=\underset{j}{\textrm{argmax}}\;p_j$ and suppose $p_i>0$, If $i$ is not a strict maximum, profits can be made through undercutting, so we assume that it is a strict maximum. Then $\mathbb{E}\left[\sum_{t=0}^T\left|\mathcal{X}_i^t(p)\right|\right]=0$, and thus a profitable deviation exists to $p_i-\epsilon$ for some $\epsilon>0$. Since $p_i$ is maximal and a deviation exists if it is positive, the pure strategy equilibrium must be $p_j=0$ for all $j$.
\item Since the allocation to each firm is independent of the price vector in this case, the pure strategy equilibrium is for each firm to set the maximum price, $p_i=R$.
\end{enumerate}
\endproof
%

The following technical result is necessary for the proof of Lemma \ref{Nfirms}.
\begin{lemma}\label{M2pinc}
$\sum_{t=0}^T\left|\mathcal{X}_i^t(p)\right|$ is nonincreasing in $p_i$ and nondecreasing in $p_{-i}$.
\end{lemma}
Observe that $\sum_{t=0}^T\left|\mathcal{X}_i^t(p)\right|$ only increases or decreases when the price-wise ordering of the firms changes, because otherwise the allocation (\ref{allo1})-(\ref{allo3}) is unaffected. Since Lemma~\ref{M2pinc} holds for any realization of the imbalance sequence $B$, it implies that $\mathbb{E}\left[\sum_{t=0}^T\left|\mathcal{X}_i^t(p)\right|\right]$ is nonincreasing in $p_i$ and nondecreasing in $p_{-i}$ as well.

\proof{(Lemma \ref{M2pinc}).}
Suppose that $p_j$ is the smallest price larger than $p_i$, and consider an increase in $p_i$ to $p_i+\epsilon$, $\epsilon<0$. If $p_i+\epsilon<p_j$, then $\sum_{t=0}^T\left|\mathcal{X}_i^t(p)\right|$ does not change. Assume that  $p_i+\epsilon>p_j$. We proceed inductively.

Without loss of generality, assume $B^0\geq0$. By construction, $|S_i^1(p)-S_i^1(p_{-i},p_i+\epsilon)|= \left|\mathcal{X}_i^0(p)\right|- \left|\mathcal{X}_i^0(p_{-i},p_i+\epsilon)\right|\geq0$, which establishes the base case. Now assume that  $\sum_{t=0}^k\left|\mathcal{X}_i^{t}(p)\right| - \sum_{t=0}^k\left|\mathcal{X}_i^{t}(p_{-i},p_i+\epsilon)\right|\geq \left|S_i^{k+1}(p)-S_i^{k+1}(p_{-i},p_i+\epsilon)\right|$. Without loss of generality, assume that $B^{k+1}>0$ and thus $\mathcal{X}_i^{k+1}(p)\geq0$ and $\mathcal{X}_i^{k+1}(p_{-i},p_i+\epsilon)\geq0$. First, consider the case that $\mathcal{X}_i^{k+1}(p) -\mathcal{X}_i^{k+1}(p_{-i},p_i+\epsilon)\leq 0$. This implies $S_i^{k+1}(p)-S_i^{k+1}(p_{-i},p_i+\epsilon)\geq0$ because $S_i^{k+1}(p)=S_i$. By assumption, we then have
\begin{eqnarray*}
\sum_{t=0}^{k+1}\left|\mathcal{X}_i^{t}(p)\right| - \sum_{t=0}^{k+1}\left|\mathcal{X}_i^{t}(p_{-i},p_i+\epsilon)\right| &\geq&\alpha_i \left|S_i^{k+1}(p)-S_i^{k+1}(p_{-i},p_i+\epsilon)\right| \\
&&+ \left|\mathcal{X}_i^{k+1}(p)\right| -\left|\mathcal{X}_i^{k+1}(p_{-i},p_i+\epsilon)\right| \\
&=&\alpha_i S_i^{k+1}(p)-\alpha_iS_i^{k+1}(p_{-i},p_i+\epsilon)\\
&& + \mathcal{X}_i^{k+1}(p) - \mathcal{X}_i^{k+1}(p_{-i},p_i+\epsilon) \\
&=&  \left|S_i^{k+2}(p)-S_i^{k+2}(p_{-i},p_i+\epsilon)\right|,
\end{eqnarray*}
where the last line is due to (\ref{Sevo}) and the fact that $\left|\mathcal{X}_i^{k+1}(p_{-i},p_i+\epsilon)\right|-\left|\mathcal{X}_i^{k+1}(p)\right|\leq \alpha_i\left|S_i^{k+1}(p)-S_i^{k+1}(p_{-i},p_i+\epsilon)\right|$ due to the capacity limit.

Now assume that that $\mathcal{X}_i^{k+1}(p) - \mathcal{X}_i^{k+1}(p_{-i},p_i+\epsilon)\geq 0$. Then 
\begin{eqnarray*}
\sum_{t=0}^{k+1}\left|\mathcal{X}_i^{t}(p)\right| - \sum_{t=0}^{k+1}\left|\mathcal{X}_i^{t}(p_{-i},p_i+\epsilon)\right| &\geq& \alpha_i\left|S_i^{k+1}(p)-S_i^{k+1}(p_{-i},p_i+\epsilon)\right| \\
&&+ \left|\mathcal{X}_i^{k+1}(p)\right| -\left|\mathcal{X}_i^{k+1}(p_{-i},p_i+\epsilon)\right| \\
&=&\alpha_i \left|S_i^{k+1}(p)-S_i^{k+1}(p_{-i},p_i+\epsilon)\right| \\
&&+ \left|\mathcal{X}_i^{k+1}(p) -\mathcal{X}_i^{k+1}(p_{-i},p_i+\epsilon)\right| \\
&\geq & \left|\alpha_iS_i^{k+1}(p) + \mathcal{X}_i^{k+1}(p) \right.\\
&&\left.-\alpha_iS_i^{k+1}(p_{-i},p_i+\epsilon) -\mathcal{X}_i^{k+1}(p_{-i},p_i+\epsilon)\right| \\
&=&  \left|S_i^{k+2}(p)-S_i^{k+2}(p_{-i},p_i+\epsilon)\right|,
\end{eqnarray*}
where the third line is due to the triangle inequality. Since $\left|S_i^{k+2}(p)-S_i^{k+2}(p_{-i},p_i+\epsilon)\right| \geq 0$, we have by induction that $\sum_{t=0}^{k}\left|\mathcal{X}_i^{t}(p)\right| - \sum_{t=0}^{k}\left|\mathcal{X}_i^{t}(p_{-i},p_i+\epsilon)\right|\geq0$ for all $k$, which establishes the desired result that $\sum_{t=0}^T\left|\mathcal{X}_i^t(p)\right|$ is nonincreasing in $p_i$.

$\sum_{t=0}^T\left|\mathcal{X}_i^t(p)\right|$ can be shown to be nondecreasing in $p_{-i}$ by reducing some $p_j$, $j\neq i$, by $\epsilon>0$ and repeating the above argument.
\endproof


\proof{(Lemma \ref{Nfirms}).}
The proof is a straightforward extension of the approach of \cite{Acemoglu2009price} to the case of random demand over multiple time periods. We first state several standard facts from game theory (see, e.g., \cite{osborne1994course}). Let $U_i$ and $L_i$ be the upper and lower boundaries of the support of $\mu_i$. By the definition of mixed-strategy equilibrium, there exists a $\pi_i^*$ and a subset $P_i\subseteq [L_i,U_i]$, $\mu_i(P_i)=1$ for which
\begin{eqnarray}
\pi_i(p_i,\mu_{-i})\leq \pi_i^*&&\forall\; p_i\in [U_i,L_i]\\
\pi_i(p_i,\mu_{-i}) = \pi_i^*&&\forall\; p_i\in P_i\label{payUL}
\end{eqnarray}
This means that there may be a few locations, e.g., a finite number of discrete points, inside $[U_i,L_i]$ but not $P_i$ at which $\pi_i(p_i,\mu_{-i})\leq \pi_i^*$. The payoff $\pi_i(p_i,\mu_{-i})$ is continuous at $p_i$ if $\mu_{-i}$ and has no atom there ($\mu_i$ has an \textit{atom} at $p_i\in [L_i,U_i]$ if $\textrm{Prob}(p_i)=a>0$, or, equivalently, $\mu_i(x)=a\delta(x-p_i)$, where $\delta(x)$ is the Dirac delta function). Therefore, $\pi_i(p_i,\mu_{-i})=\pi_i^*$ if $\mu_{-i}$ has no atom at $p_i$.

We now prove each point of the lemma sequentially.
\begin{enumerate}
\item First suppose that $\mu_i(p)=0$ for all $i$ and all $p_i\in [\underline{p},\overline{p}]$, $L_j<\underline{p}<\overline{p}<U_j$ for at least one firm $j$. Then from $\underline{p}$ to $p\in (\underline{p},\overline{p})$ is a profitable deviation for any firm with probability mass below $\underline{p}$. Now suppose that only one firm $i$ has $\mu_i(p)>0$ for $p_i\in [\underline{p},\overline{p}]$. Then the distribution
\begin{displaymath}
\mu'_i(p)=\left\{\begin{array}{ll} \mu_i(p) &\textrm{if } p_i<\underline{p} \\
0 &\textrm{if } \underline{p}\leq p_i<\overline{p}\\
\mu\left( [\underline{p},\overline{p}] \right) &\textrm{if } p_i=\overline{p}\\
\mu_i(p) &\textrm{if } p_i>\overline{p}
 \end{array}\right.
\end{displaymath}
is a profitable deviation for $i$; essentially, all of firm $i$'s probability mass in $[\underline{p},\overline{p}]$ has been shifted to $\overline{p}$. Thus at least two firms $i$ must have $\mu_i(p)>0$ for all $j$ with $p_i\in [L_j, U_j]$.

\item First we show that no two firms may have an atom at the same location. Suppose multiple firms have an atom at $p'$. Then with positive probability all such firms set $p_i=p'$. For a given firm $i$, there exists an $\epsilon>0$ small enough that $\pi_i(p'-\epsilon,\mu_{-i})>\pi_i(p',\mu_{-i})$, a profitable deviation for firm $i$.

We now show that no firm can have an atom in $[L,U)$. Suppose firm $i$ has an atom at $p$. If $p\notin [L_j,U_j)$ for all $j\neq i$, then $p+\epsilon$, $\epsilon>0$ is a profitable deviation for firm $i$. Now assume firm $i$ has an atom of mass $a\in(0,1]$ at $p'\in (L_j,U_j)$ for some $j\neq i$, and let $\hat{\mu}$ denote $\mu$ with the atom subtracted off. Consider the difference between firm $j$'s profits at $p'-\epsilon$ and $p'+\epsilon$:
\begin{eqnarray*}
&&\pi_j\left(p'-\epsilon,\mu_{-j}\right)-\pi_j\left(p'+\epsilon,\mu_{-j}\right)=\\
&&\quad\left(p'-\epsilon\right)\left[\int_{p_{-j}}\left(\prod_{k\neq j}\hat{\mu}_k(p_k)\right)\mathbb{E}\left[\sum_{t=0}^T\left|\mathcal{X}_j^t\left(p'-\epsilon,p_{-j}\right)\right|\right] dp\right.\\
&&\quad+\left.a\int_{p_{-i,j}}\left(\prod_{k\neq i,j}\hat{\mu}_k(p_k)\right)\mathbb{E}\left[\sum_{t=0}^T\left|\mathcal{X}_j^t\left(p'-\epsilon,p',p_{-i,j}\right)\right|\right] dp\right]\\
&&\quad-\left(p'+\epsilon\right)\left[\int_{p_{-j}}\left(\prod_{k\neq j}\hat{\mu}_k(p_k)\right)\mathbb{E}\left[\sum_{t=0}^T\left|\mathcal{X}_j^t\left(p'+\epsilon,p_{-j}\right)\right|\right] dp\right.\\
&&\quad+\left.a\int_{p_{-i,j}}\left(\prod_{k\neq i,j}\hat{\mu}_k(p_k)\right)\mathbb{E}\left[\sum_{t=0}^T\left|\mathcal{X}_j^t\left(p'+\epsilon,p',p_{-i,j}\right)\right|\right] dp\right]\\
\end{eqnarray*}
Letting $\epsilon$ tend to zero, the first and third terms cancel, while the sum of the remaining terms remains positive because $\sum_{t=0}^T\left|\mathcal{X}_j^t(p)\right|$ (by Lemma \ref{M2pinc}) and hence $\mathbb{E}\left[\sum_{t=0}^T\left|\mathcal{X}_j^t(p)\right|\right]$ are nonincreasing in $p_j$. Since firm $j$ has no atom at $p'-\epsilon$ or $p'+\epsilon$ for some $\epsilon$, its strategy is continuous at $p'+\epsilon$ and $\pi_j\left(p'+\epsilon,\mu_{-j}\right)=\pi_j^*$. But $p'-\epsilon$ is a profitable deviation from $p'+\epsilon$ for firm $j$, establishing that an atom can only exist at $U$.

\item Suppose $U<R$. First  consider the case that a firm, $i$, has an atom at $U$. Then $\pi_i(R,\mu_{-i})>\pi_i(U,\mu_{-i})$, and a profitable deviation exists. Now suppose no firm has an atom at $U$. Then similarly for any firm $i$ with upper support $U$, $\pi_i(R,\mu_{-i})>\pi_i(U,\mu_{-i})$, establishing the existence of a profitable deviation.
\end{enumerate}
\endproof

\proof{(Theorem \ref{prop:mu}).}
Let $\Upsilon_i$ be the cumulative distribution of firm $i$'s mixed-strategy. Since, by Lemma~\ref{Nfirms}, neither firm has an atom in $[L,R)$, we have that 
\begin{eqnarray*}
\pi_1^*&=&x\left(\int_L^{x}\mu_2(p_2)\mathbb{E}\left[\sum_{t=0}^T\left|\mathcal{X}_1^t(x,p_2)\right|\right] dp_2 +  \int_{x}^{R}\mu_2(p_2)\mathbb{E}\left[\sum_{t=0}^T\left|\mathcal{X}_1^t(x,p_2)\right|\right]dp_2 \right)\\
&=&x\left(\overline{\mathcal{X}}_1\int_L^{x}\mu_2(p_2)dp_2+\underline{\mathcal{X}}_1\int_{x}^{R}\mu_2(p_2)dp_2  \right)\\
&=&x\left(\overline{\mathcal{X}}_1\Upsilon_2(x)+\underline{\mathcal{X}}_1(1-\Upsilon_2(x))  \right).
\end{eqnarray*}
Solving for $\Upsilon_2(x)$ over $[L,R)$, we have
\begin{eqnarray*}
\Upsilon_2(x)&=&\frac{\underline{\mathcal{X}}_1}{\underline{\mathcal{X}}_1-\overline{\mathcal{X}}_1}
-\frac{\pi_1^*}{\left(\underline{\mathcal{X}}_1-\overline{\mathcal{X}}_1\right)x}.
\end{eqnarray*}
By the same argument, we similarly have
\begin{eqnarray*}
\Upsilon_1(x)&=&\frac{\underline{\mathcal{X}}_2}{\underline{\mathcal{X}}_2-\overline{\mathcal{X}}_2}
-\frac{\pi_2^*}{\left(\underline{\mathcal{X}}_2-\overline{\mathcal{X}}_2\right)x},
\end{eqnarray*}
also over $[L,R)$.

From Lemma~\ref{Nfirms}, only one firm can have an atom at $R$. If $S_1=S_2$, they are interchangeable. Assume now that $S_1>S_2$ and that $\underline{\mathcal{X}}_1\geq\underline{\mathcal{X}}_2$ and $\overline{\mathcal{X}}_1\geq\overline{\mathcal{X}}_2$. Suppose for the sake of contradiction that firm two, which has smaller capacity, has an atom at $R$. Then, since firm one cannot have an atom at $R$,
\begin{eqnarray*}
\pi_2^*&=&R\overline{\mathcal{X}}_2\\
\pi_1^*&=&\frac{R\overline{\mathcal{X}}_2\underline{\mathcal{X}}_1}{\underline{\mathcal{X}}_2}
\end{eqnarray*}
Subbing $\overline{\pi}(C,R,S)$ into the above expression for $\Upsilon_2(R)$, we have 
\begin{eqnarray*}
\Upsilon_2(R)&=&\frac{\underline{\mathcal{X}}_1}{\underline{\mathcal{X}}_1-\overline{\mathcal{X}}_1}
-\frac{\overline{\mathcal{X}}_2\underline{\mathcal{X}}_1}{\underline{\mathcal{X}}_2\left(\underline{\mathcal{X}}_1-\overline{\mathcal{X}}_1\right)}\\
&>&\frac{\underline{\mathcal{X}}_1}{\underline{\mathcal{X}}_1-\overline{\mathcal{X}}_1}
-\frac{\overline{\mathcal{X}}_2\underline{\mathcal{X}}_1}{\underline{\mathcal{X}}_1\left(\underline{\mathcal{X}}_1-\overline{\mathcal{X}}_1\right)} \quad (\textrm{because} \; \underline{\mathcal{X}}_1>\underline{\mathcal{X}}_2)\\
&=& \frac{\underline{\mathcal{X}}_1-\overline{\mathcal{X}}_2}{\underline{\mathcal{X}}_1-\overline{\mathcal{X}}_1} \\
&>&1 \quad (\textrm{because} \; \overline{\mathcal{X}}_1>\overline{\mathcal{X}}_2).
\end{eqnarray*}

This contradicts the assumption that $\mu_2$ has an atom at $R$. Therefore, the equilibrium payoffs are 
\begin{eqnarray*}
\pi_1^*&=&R\overline{\mathcal{X}}_1\\
\pi_2^*&=&\frac{R\overline{\mathcal{X}}_1\underline{\mathcal{X}}_2}{\underline{\mathcal{X}}_1}
\end{eqnarray*}
Now setting $\Upsilon_2(L)=0$, we have that $L=R\overline{\mathcal{X}}_1/\underline{\mathcal{X}}_1$. The atom at $R$ in the large firm's strategy can be shown to be
\begin{displaymath}
\frac{\underline{\mathcal{X}}_2\overline{\mathcal{X}}_1-\overline{\mathcal{X}}_2\underline{\mathcal{X}}_1}{\underline{\mathcal{X}}_1\left(\underline{\mathcal{X}}_2-\overline{\mathcal{X}}_2\right)}
\end{displaymath}

 The mixed strategies, can be obtained by substituting the above into $\Upsilon_i(x)$ and differentiating.
\endproof


\proof{(Lemma \ref{capexist}).}
The second derivative of $\overline{\pi}(S_i,S_{-i})-\gamma_iS_i$ with respect to $S_i$ is $-Rf(S_{i}+S_{-i})$, and hence it is strictly concave because $f$ is positive. Since $\underline{\pi}(S_{-i},S_i)$ is also differentiable, it suffices for quasiconcavity to show that its first derivative with respect to $S_i$ is initially positive and crosses zero exactly once \cite{boyd2004convex}. Let $\overline{S}_i=\int_0^{S_i}Bf(B)dB+S_i(1-F(S_i))$. The first derivative is given by
\begin{eqnarray}
\frac{d\underline{\pi}(S_{-i},S_i)}{dS_i} & = & \frac{1}{\overline{S}_{-i}}\left[(1-F(S_i))\overline{\pi}(S_{-i},S_i) + R\left(F(S_{i})-F(S_i+S_{-i})\right)\overline{S}_{i}\right]\label{dpi2ds2p}
\end{eqnarray}
where $\overline{S}_i=\int_0^{\infty}\min(S_i,D)f(D)dD$. Dividing through by $\overline{S}_i$, define
\begin{eqnarray*}
M(S_i) &=& \frac{(1-F(S_i))\overline{\pi}(S_{-i},S_i)}{\overline{S}_i}\\
 N(S_i) &=& R\left(F(S_i)-F(S_i+S_{-i})\right)
\end{eqnarray*}
$\underline{\pi}(S_{-i},S_i)$ is equal to zero at $S_i=0$ and $S_i=\infty$ but is not constantly zero. Therefore, by the mean value theorem, (\ref{dpi2ds2p}) is zero at least once in between, which implies that $M(S_i) +N(S_i)=0$ at any such point.

Observe that $M(S_i)$ is always positive and $N(S_i)$ always negative. One may straightforwardly differentiate to see that the magnitude of $(1-F(S_i))$ shrinks faster than that of $(F(S_i)-F(S_i+S_{-i}))$ as $S_i$ increases; since the fraction $\frac{\overline{\pi}(S_{-i},S_i)}{\overline{S}_i}$ also approaches zero with $S_i$, we may conclude that the magnitude of $M(S_i)$ shrinks faster than that of $N(S_i)$. Since the magnitudes are identical at any point $S_i'$ where $M(S_i') +N(S_i')=0$, then $M(S_i'') +N(S_i'')<0$ for any $S_i''>S_i'$. Therefore (\ref{dpi2ds2p}) is zero at only one finite value, henceforth denoted $S_i'$, establishing the quasiconcavity of $\underline{\pi}(S_{-i},S_i)$ with respect to $S_i$. 

Because the magnitude of $M(S_i)$ shrinks faster than that of $N(S_i)$, $\frac{d\underline{\pi}}{dS_2}(S_{-i},S_i)$ is strictly decreasing in $S_i$ prior to crossing zero. Since $-\gamma_i$ is negative and constant, $\frac{d\underline{\pi}(S_{-i},S_i)}{dS_i}-\gamma_i$ is also strictly decreasing in $S_i$ before crossing zero and then remains negative, implying quasiconcavity of $\underline{\pi}(S_{-i},S_i)-\gamma_{-i}S_{-i}$ with respect to $S_i$.
\endproof

\proof{(Theorem \ref{M2capeqs}).}
We prove that $\hat{S}^i$ are the only possible pure-strategy equilibria by showing that an equilibrium can only exist where $\psi_i$ has zero slope in $S_i$. This is equivalent to showing that no `ridges' exist in $\psi_i$. A sufficient condition is
\begin{equation}
\left. \frac{d\underline{\pi}\left(S_{-i},S_i\right)}{dS_i}\right|_{S_i=S_{-i}}\leq \left.\frac{d\overline{\pi}\left(S_i,S_{-i}\right)}{dS_i}\right|_{S_i=S_{-i}}.
\end{equation}
We have that
\begin{eqnarray*}
\left.\frac{d\underline{\pi}\left(S_{-i},S_i\right) }{dS_i}\right|_{S_{-i}=S_i}&=& \left(1-F\left(S_i\right)\right)\frac{\overline{\pi}\left(S_i,S_i\right)}{\overline{S}_i}+R\left(F\left(S_i\right)-F\left(2S_i\right)\right)\\
&\leq & R\left(1-F\left(S_i\right)\right)+R\left(F\left(S_i\right)-F\left(2S_i\right)\right)\\
&= & R\left(1-F\left(2S_i\right)\right)\\
&= & \left.\frac{d\overline{\pi}\left(S_i,S_{-i}\right)}{dS_i}\right|_{S_{-i}=S_i},
\end{eqnarray*}
establishing the claim.
\endproof

\bibliographystyle{plain}
\bibliography{MainBib}

\begin{thebibliography}{10}

\bibitem{Acemoglu2009price}
D.~Acemoglu, K.~Bimpikis, and A.~Ozdaglar.
\newblock Price and capacity competition.
\newblock {\em Games and Economic Behavior}, 66(1):1 -- 26, 2009.

\bibitem{Acemoglu2007cong}
D.~Acemoglu and A.~Ozdaglar.
\newblock Competition and efficiency in congested markets.
\newblock {\em Math. Oper. Res.}, 32:1--31, February 2007.

\bibitem{NYISO2011}
D.~Allen, C.~Brown, J.~Hickey, V.~Le, and R.~Safuto.
\newblock Energy storage in the {New York} electricity markets.
\newblock New York Independent System Operator, Mar. 2010.

\bibitem{alvarez04assessment}
C.~Alvarez, A.~A.~Gabald\'{o}n, and A.~Molina.
\newblock Assessment and simulation of the responsive demand potential in
  end-user facilities: Application to a university customer.
\newblock {\em IEEE Transactions on Power Systems}, 19(2):1223--1231, 2004.

\bibitem{baldick2004lin}
R.~Baldick, R.~Grant, and E.~Kahn.
\newblock Theory and application of linear supply function equilibrium in
  electricity markets.
\newblock {\em Journal of Regulatory Economics}, 25:143--167, 2004.

\bibitem{infield2004storage}
J.P. Barton and D.G. Infield.
\newblock Energy storage and its use with intermittent renewable energy.
\newblock {\em IEEE Transactions on Energy Conversion}, 19(2):441 -- 448, Jun.
  2004.

\bibitem{beaconenergy2011}
{Beacon Power Corp.}
\newblock Energy storage: Regulation issues.
\newblock {ERCOT} Emerging Technologies Working Group, 2011.

\bibitem{bertrand1883}
J.~Bertrand.
\newblock Theorie mathematique de la richesse sociale.
\newblock {\em Journaldes Savants}, pages 499--508, 1883.

\bibitem{billinton1996power}
Roy Billinton and Ronald~N. Allan.
\newblock {\em Reliability evaluation of power systems}.
\newblock Springer, second edition, 1996.

\bibitem{Bolle1992supply}
F.~Bolle.
\newblock Supply function equilibria and the danger of tacit collusion: The
  case of spot markets for electricity.
\newblock {\em Energy Economics}, 14(2):94 -- 102, 1992.

\bibitem{boyd2004convex}
S.~Boyd and L.~Vandenberghe.
\newblock {\em Convex Optimization}.
\newblock Cambridge University Press, New York, NY, USA, 2004.

\bibitem{oren1994bidder}
James~B. Bushnell and Shmuel~S. Oren.
\newblock Bidder cost revelation in electric power auctions.
\newblock {\em Journal of Regulatory Economics}, 6(1):5--26, 1994.

\bibitem{caiso_bpm_2014}
CAISO.
\newblock Business practice manual for market operations, version 39.
\newblock Technical report, California Independent System Operator Business
  Practice Manuals Library, 2014.

\bibitem{REMdraft}
{California ISO}.
\newblock Regulation energy management draft final proposal, January 2011.

\bibitem{caiso_pay_2012}
{California ISO}.
\newblock Pay for performance regulation: Draft final proposal, Mar. 2012.

\bibitem{callaway11EL}
{D.S.} Callaway and {I.A.} Hiskens.
\newblock Achieving controllability of electric loads.
\newblock {\em Proceedings of the {IEEE}}, 99(1):184--199, January 2011.

\bibitem{carrasco2006pes}
J.M. Carrasco, L.G. Franquelo, J.T. Bialasiewicz, E.~Galvan, R.C.P. Guisado,
  Ma.A.M. Prats, J.I. Leon, and N.~Moreno-Alfonso.
\newblock Power-electronic systems for the grid integration of renewable energy
  sources: A survey.
\newblock {\em IEEE Transactions on Industrial Electronics}, 53(4):1002 --1016,
  Jun. 2006.

\bibitem{Castillo2014survey}
Anya Castillo and Dennice~F. Gayme.
\newblock Grid-scale energy storage applications in renewable energy
  integration: A survey.
\newblock {\em Energy Conversion and Management}, 87(0):885 -- 894, 2014.

\bibitem{wilson2002multi}
Hung-Po Chao and Robert Wilson.
\newblock Multi-dimensional procurement auctions for power reserves: Robust
  incentive-compatible scoring and settlement rules.
\newblock {\em Journal of Regulatory Economics}, 22(2):161--183, 2002.

\bibitem{Cramton2005cap}
Peter Cramton and Steven Stoft.
\newblock A capacity market that makes sense.
\newblock {\em The Electricity Journal}, 18(7):43 -- 54, 2005.

\bibitem{maskin1986theory}
P.~Dasgupta and E.~Maskin.
\newblock The existence of equilibrium in discontinuous economic games, {I}:
  {T}heory.
\newblock {\em The Review of Economic Studies}, 53(1):1--26, 1986.

\bibitem{hobbs2002supply}
C.J. Day, B.F. Hobbs, and {J.S.} Pang.
\newblock Oligopolistic competition in power networks: a conjectured supply
  function approach.
\newblock {\em IEEE Transactions on Power Systems}, 17(3):597 -- 607, Aug.
  2002.

\bibitem{Dunn2011choices}
Bruce Dunn, Haresh Kamath, and Jean-Marie Tarascon.
\newblock Electrical energy storage for the grid: A battery of choices.
\newblock {\em Science}, 334(6058):928--935, 2011.

\bibitem{edgeworth1925papers}
F.Y. Edgeworth.
\newblock The pure theory of monopoly.
\newblock In {\em Papers relating to political economy}, volume~1, pages
  111--142. Macmillan and Co., Ltd., 1925.

\bibitem{Fabra2006auctions}
Natalia Fabra, Nils-Henrik von~der Fehr, and David Harbord.
\newblock Designing electricity auctions.
\newblock {\em The RAND Journal of Economics}, 37(1):23--46, 2006.

\bibitem{jpmorgan2013ferc}
{FERC}.
\newblock {FERC}, {JP Morgan} unit agree to \$410 million in penalties,
  disgorgement to ratepayers, July 2013.

\bibitem{first_dinorwig_2012}
{First Hydro Company}.
\newblock Dinorwig power station, 2009.

\bibitem{fudenberg1991GT}
D.~Fudenberg and J.~Tirole.
\newblock {\em {Game Theory}}.
\newblock MIT Press, 1991.

\bibitem{Glicksberg1952Nash}
I.~L. Glicksberg.
\newblock A further generalization of the {K}akutani fixed point theorem, with
  application to nash equilibrium points.
\newblock {\em Proceedings of the American Mathematical Society},
  3(1):170--174, 1952.

\bibitem{green1992british}
{R.J.} Green and {D.M.} Newbery.
\newblock Competition in the {B}ritish electricity spot market.
\newblock {\em Journal of Political Economy}, 100(5):929--953, 1992.

\bibitem{halamay2011reserve}
D.A. Halamay, T.K.A. Brekken, A.~Simmons, and S.~McArthur.
\newblock Reserve requirement impacts of large-scale integration of wind,
  solar, and ocean wave power generation.
\newblock {\em IEEE Transactions on Sustainable Energy}, 2(3):321--328, Jul.
  2011.

\bibitem{he2011novel}
Xian He, Erik Delarue, William D'haeseleer, and Jean-Michel Glachant.
\newblock A novel business model for aggregating the values of electricity
  storage.
\newblock {\em Energy Policy}, 39(3):1575--1585, 2011.

\bibitem{Ibrahim2008char}
H.~Ibrahim, A.~Ilinca, and J.~Perron.
\newblock Energy storage systems - characteristics and comparisons.
\newblock {\em Renewable and Sustainable Energy Reviews}, 12(5):1221 -- 1250,
  2008.

\bibitem{Joskow2001crisis}
{P.L.} Joskow.
\newblock California's electricity crisis.
\newblock {\em Oxford Review of Economic Policy}, 17(3):365--388, 2001.

\bibitem{Kahn2001dil}
Alfred~E. Kahn, Peter~C. Cramton, Robert~H. Porter, and Richard~D. Tabors.
\newblock Uniform pricing or pay-as-bid pricing: A dilemma for {C}alifornia and
  beyond.
\newblock {\em The Electricity Journal}, 14(6):70 -- 79, 2001.

\bibitem{klemperer1989supply}
{P.D.} Klemperer and {M.A.} Meyer.
\newblock Supply function equilibria in oligopoly under uncertainty.
\newblock {\em Econometrica}, 57(6):1243--1277, 1989.

\bibitem{kremer2004role}
Ilan Kremer and Kjell~G. Nyborg.
\newblock Divisible-good auctions: The role of allocation rules.
\newblock {\em The RAND Journal of Economics}, 35(1):147--159, 2004.

\bibitem{kreps1983cournot}
{D.M.} Kreps and {J.A.} Scheinkman.
\newblock Quantity precommitment and {B}ertrand competition yield {C}ournot
  outcomes.
\newblock {\em The Bell Journal of Economics}, 14(2):326--337, 1983.

\bibitem{kundur1994stab}
Prabha Kundur.
\newblock {\em Power system stability and control}.
\newblock McGraw-Hill Professional, 1994.

\bibitem{makarov09}
Y.~V Makarov, C.~Loutan, J.~Ma, and P.~De~Mello.
\newblock Operational impacts of wind generation on {California} power systems.
\newblock {\em {IEEE} Transactions on Power Systems}, 24(2):1039--1050, 2009.

\bibitem{mathieu_energy_2013}
{J.L.} Mathieu, M.~Kamgarpour, J.~Lygeros, and {D.S.} Callaway.
\newblock Energy arbitrage with thermostatically controlled loads.
\newblock In {\em Proceedings of the European Control Conference}, pages
  2519--2526, Z\"{u}rich, Switzerland, 2013.

\bibitem{taylor2013FlexCDC}
A.~Nayyar, J.~A. Taylor, A.~Subramanian, D.~S. Callaway, and K.~Poolla.
\newblock Aggregate flexibility of collections of loads.
\newblock In {\em Decision and Control (CDC), IEEE 52nd Annual Conference on},
  pages 5600--5607, Dec. 2013.
\newblock Invited.

\bibitem{oren2000cap}
Shmuel Oren.
\newblock Capacity payments and supply adequacy in competitive electricity
  markets.
\newblock In {\em VII SEPOPE}, pages 1--8, May 2000.

\bibitem{osborne1994course}
M.J. Osborne and A.~Rubinstein.
\newblock {\em A course in game theory}.
\newblock MIT Press, 1994.

\bibitem{pakes1994nash}
Ariel Pakes and Paul McGuire.
\newblock Computing {M}arkov-perfect {N}ash equilibria: Numerical implications
  of a dynamic differentiated product model.
\newblock {\em The {RAND} Journal of Economics}, 25(4):555--589, 1994.

\bibitem{oren2011wind}
A.~Papavasiliou, S.S. Oren, and R.P. O'Neill.
\newblock Reserve requirements for wind power integration: A scenario-based
  stochastic programming framework.
\newblock {\em Power Systems, IEEE Transactions on}, 26(4):2197--2206, Nov.
  2011.

\bibitem{peterson10economics}
S.~B. Peterson, J.~F. Whitacre, and J.~Apt.
\newblock The economics of using plug-in hybrid electric vehicle battery packs
  for grid storage.
\newblock {\em Journal of Power Sources}, 195:2377--2384, 2010.

\bibitem{Sioshansi2014when}
Ramteen Sioshansi.
\newblock When energy storage reduces social welfare.
\newblock {\em Energy Economics}, 41(0):106 -- 116, 2014.

\bibitem{sioshansia2012market}
Ramteen Sioshansi, Paul Denholm, and Thomas Jenkin.
\newblock Market and policy barriers to deployment of energy storage.
\newblock {\em Economics of Energy and Environmental Policy Journal}, 1(2):47,
  2012.

\bibitem{su09quantifying}
C.~L. Su and D.~Kirschen.
\newblock Quantifying the effect of demand response on electricity markets.
\newblock {\em IEEE Transactions on Power Systems}, 24(3):1199--1207, 2009.

\bibitem{taylor2011store}
J.~A. Taylor, D.~S. Callaway, and K.~Poolla.
\newblock Competitive energy storage in the presence of renewables.
\newblock {\em Submitted, IEEE Trans. on Power Systems}, 2011.

\bibitem{vittal_impact_2009}
V.~Vittal, J.~{McCalley}, V.~Ajjarapu, and U.V. Shanbhag.
\newblock Impact of increased {DFIG} wind penetration on power systems and
  markets.
\newblock Technical Report {PSERC} 09-10, Power Systems Engineering Research
  Center, 2009.

\bibitem{vives2001oligopoly}
X.~Vives.
\newblock {\em Oligopoly pricing: old ideas and new tools}.
\newblock MIT Press, 2001.

\bibitem{WoodWoll2}
A.~J. Wood and B.~F. Wollenberg.
\newblock {\em Power generation, operation, and control}.
\newblock Wiley, 3rd edition, 2013.

\bibitem{Litvinov2006ex}
T.~Zheng and E.~Litvinov.
\newblock Ex post pricing in the co-optimized energy and reserve market.
\newblock {\em IEEE Transactions on Power Systems}, 21(4):1528 --1538, Nov.
  2006.

\end{thebibliography}

\end{document}